# Effects of Partial Hydrogenation on the Structure and Electronic Properties of BNNTs

Lena Kalikhman-Razvozov, Roza Yusupov, and Oded Hod

Department of Chemical Physics, School of Chemistry, The Raymond and Beverly Sackler Faculty of Exact Sciences, Tel-Aviv University, Tel-Aviv 69978, Israel

**Abstract**

The effects of partial hydrogenation on the structure and electronic properties of boron nitride nanotubes are investigated via density functional theory calculations. We find that the structure of the nanotube may considerably deform depending on the exact locations of the hydrogen atoms adsorption. Furthermore, depending on the tube identity, diameter, and adsorbate density it is found that the bandgap can be gradually reduced as a function of the relative position of the hydrogen atoms and in some cases change its character from indirect to direct and vice versa upon hydrogen adsorption. Our findings indicate that partial hydrogenation may prove as a valuable and experimentally achievable route for controlling the electronic properties of boron nitride nanotubes.

**Introduction**

The (meta)stability and electronic properties of boron nitride nanotubes (BNNT) were first discussed in a seminal theoretical study by Cohen et al. in 1994.[1-2] This was followed by their subsequent synthesis in 1995 [3]. Since then these systems have attracted considerable scientific attention [4] due to their unique mechanical properties, high thermal conductivity, stability, and resistance to oxidation [5-8]. Unlike the case of carbon nanotubes, where the electronic properties strongly depend on their exact morphology [9] the bandgap of BNNTs results from the inherent electronegativity differences between the boron and nitrogen atoms leading to $\pi$ electron localization around the nitrogen sites. Therefore, their measured bandgap of 5.5 eV is found to be practically independent of chirality, diameter, and number of walls [2, 10]. This high and robust bandgap strongly limits the applicability of BNNTs as potential active components in nano-electronic devices and it is therefore desirable to find practical routes for reducing their bandgap.

To this end, several schemes have been proposed to gain control over the electronic properties of BNNTs including mechanical deformations [11-14], application of external fields [15-19], structural modifications such as Stone-Wales defects [19-21], chemical doping via substitution with different atomic constituents [22-25], and chemi- and physi-sorption of different atomic and molecular entities [15-16, 26-35]. One of the promising routes considered is the chemi-sorption of atomic hydrogen on the surface of hexagonal-boron nitride and BNNTs which is important not only for controlling their electronic properties but also from the perspective of hydrogen storage.[31-33] In a recent study by Y. Wang [33] it was shown that full and semi-full hydrogenation of an infinite hexagonal boron-nitride (*h*-BN) sheet transforms the otherwise insulating system into a direct bandgap semiconductor or a metal, respectively. Interestingly, when considering similar hydrogenation schemes for tubular BN structures Tanskanen et al. found that the systems remain insulating [36]. The effects of lower hydrogen coverage on the magnetic properties of BNNTs have been further explored finding that hydrogenation induced magnetization may occur only at low hydrogen coverage [37].

In the present paper, we aim to better understand the mechanisms underlying the effects of partial hydrogenation on the structural and electronic properties of BNNTs and to further explore scenarios of lower system coverage. To this end, we study the chemi-sorption of hydrogen atoms on several BNNTs considering various adsorption schemes and densities.

**Computational details**

All calculations reported in this paper have been performed using the GAUSSIAN 09 suite of programs [38]. Here, Bloch functions are expanded in terms of atomic Gaussian-type orbitals, and the Kohn-Sham (KS) equations of density functional theory are solved self-consistently in that basis set. All geometry optimizations and electronic structure calculations have been obtained using the double-$\zeta$ polarized 6-31G** Gaussian basis set [39]. Three different exchange-correlation functional approximations have been considered including the local density approximation (LDA) [40-41], the Perdew, Burke and Ernzerhof (PBE) version of the generalized gradient approximation (GGA) [42-43], and the screened-exchange hybrid functional developed by Heyd, Scuseria and Ernzerhof (HSE) [44-47].

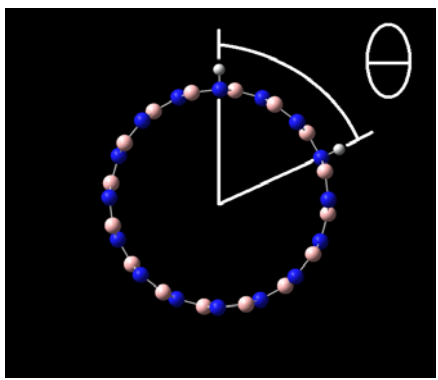

Figure 1: Definition of the angular separation between two hydrogen adsorption sites on a pristine (8,8) nanotube.

We consider four types of BNNTs including the (5,5) and (8,8) armchair BNNTs (ACBNNTs) and the (6,0) and (9,0) zigzag BNNTs (ZZBNNTs). First, we optimize the geometry of the pristine BNNTs for all functional approximations using periodic boundary conditions calculations. To mimic the dissociative adsorption of a hydrogen molecule, two hydrogen atoms are then covalently bonded to each unit-cell of the optimized structures. This also provides us with the flexibility to evaluate the effect of different spin states. The bonding scheme may be classified into three families according to the identity of the chemisorptions sites of the two hydrogen atoms in the following manner: (i) the two hydrogen atoms are located on two nitrogen atoms (NN); (ii) the two hydrogen atoms are located on two boron atoms (BB); and (iii) one hydrogen atom is located on a boron site and the second hydrogen atom is located on a nitrogen site (BN). Within each of these three families different position are marked according to the angular separation ($\theta$) of the two adsorption sites in the pristine system (see Fig. 1). Geometry optimization is then performed for all adsorption schemes of all BNNTs with the three functional approximations considered followed by an analysis of the structural and electronic properties of the system. In order to evaluate the effect of hydrogen density we reduce the hydrogen density by duplicating the unit-cells of the narrower pristine BNNTs and repeating the adsorption schemes discussed above with two hydrogen atoms per two BNNT unit-cells.

**Results and discussion**

Structural analysis

In order to study the relative structural stability of the different hydrogen adsorption schemes we plot in Fig. 2 the total energies obtained at the HSE/6-31G$^{**}$ level of theory as a function of the inter-hydrogen angular separation for the four BNNTs studied. Results obtained using the LDA and PBE functional approximations show similar trends and are presented in the supplementary material. For each structure, both the closed-shell singlet (full symbols) and open-shell triplet (open symbols) spin states were calculated and the energy of the more stable state is indicated in the diagram. The total energy of the most stable structure in each diagram is used as a reference value and is thus set to zero. We find that for all BNNTs studied the most stable structure is obtained when the hydrogen atoms adsorb at the smallest angular separation. This state is always found to be of closed-shell singlet nature. As the angular separation increases a monotonous decrease in stability is observed up to an angular separation of ~$90^0$. Beyond that, a relatively weak dependence of the total energy on the angular separation is observed for all nanotubes studied and adsorption schemes considered. Interestingly, for all the BB and most of the NN adsorption schemes the ground state is of triplet character. On the other hand, for all BN adsorption schemes with angular separation below $90^0$ the ground state is of closed shell singlet nature. We note that, for the higher adsorption density considered, the energy differences between the most stable configuration and the highest energy configuration are of the order of 4.3-5.6 eV/unit-cell depending on the system studied. Furthermore, the most stable adsorption scheme of the zigzag nanotubes studied is ~0.5 eV/unit-cell lower in energy than the next meta-stable configuration and for the armchair counterparts this energy difference is on the order of 2 eV/unit-cell. Similar trends though with somewhat reduced energy differences values are obtained when halving the hydrogen density (see supplementary material).

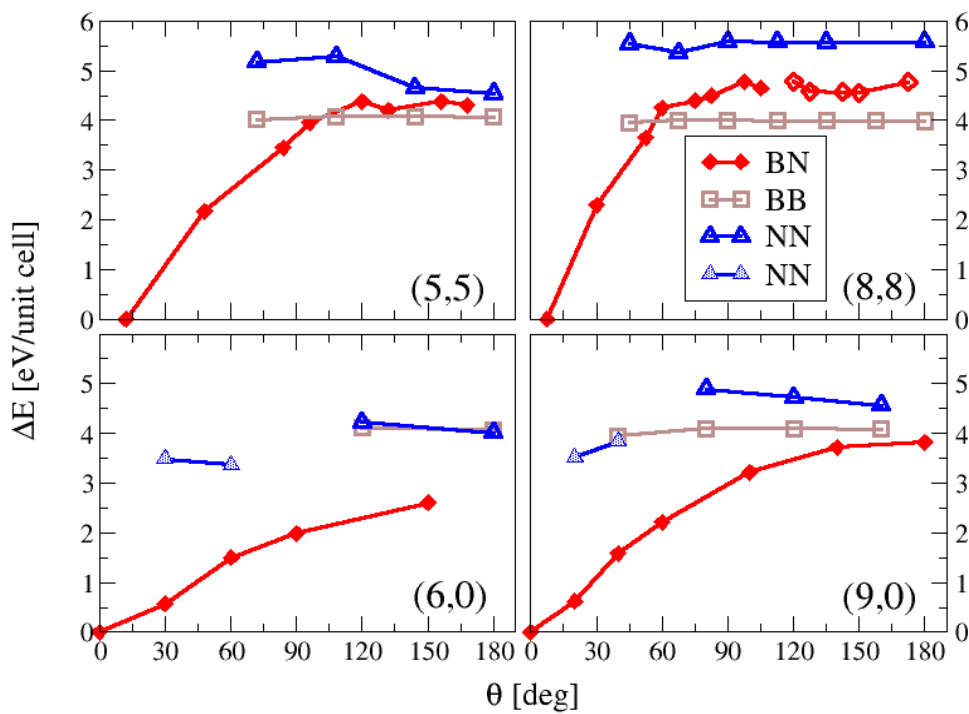

Figure 2: Relative total energies as a function of inter-hydrogen angular separation (θ) for the (5,5) and (8,8) armchair (upper panels) and (6,0) and (9,0) zigzag (lower panels) BNNTs studied obtained at the HSE/6-31G$^{**}$ level of theory. The total energy of the most stable adsorption scheme of each system is used as reference. For clarity, lines connecting values that belong to the BN (red diamonds), BB (brown squares), and NN (blue triangles) adsorption families were added. Full (open) symbols designate a closed-shell singlet (open-shell triplet) ground state. Dotted triangles represent NN adsorption schemes that form buckling deformations during geometry optimization (see Fig. 5 below and the relevant discussion).

To better understand the effect of hydrogen adsorption on the structural properties of the BNNTs, we plot in Fig. 3 axial views of the pristine and the most stable hydrogen decorated BNNTs studied. As can be seen, upon hydrogen adsorption the circular cross section characterizing the pristine systems turns into a teardrop shape. For larger hydrogen angular separations the overall strain induced by the two hydrogen atoms decreases and the structural deformation results in the formation of two cross-sectional segments of different curvatures (see upper panels of Fig. 4). Once the angular separation reaches the maximum value of 180$^o$ the two segments possess an equal curvature and an eye-shaped structure is obtained. We note that this behavior remains valid when the hydrogen adsorbate density is reduced however the overall cross-sectional deformation becomes less sever (see lower panels of Fig. 4).

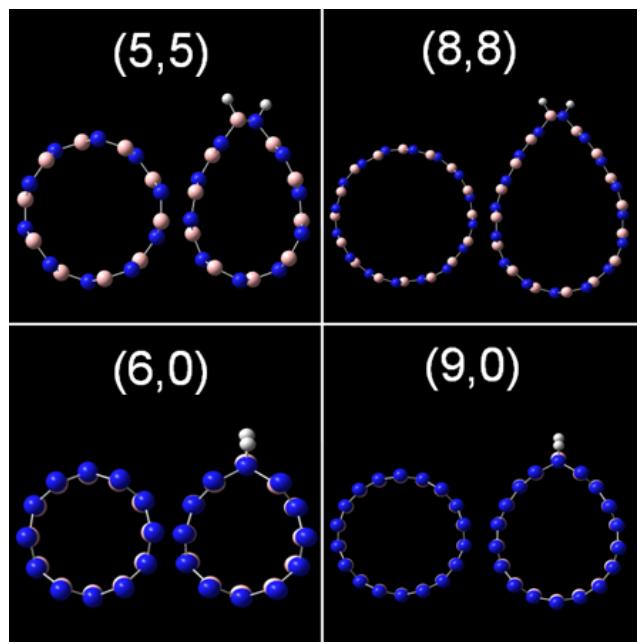

Figure 3: A schematic representation of the optimized structures of the various pristine BNNTs studied and their most stable hydrogenated counterparts obtained when adsorbing two hydrogen atoms per nanotube unit-cell. Minimum energy geometries presented were obtained at the HSE/6-31G$^{**}$ level of theory.

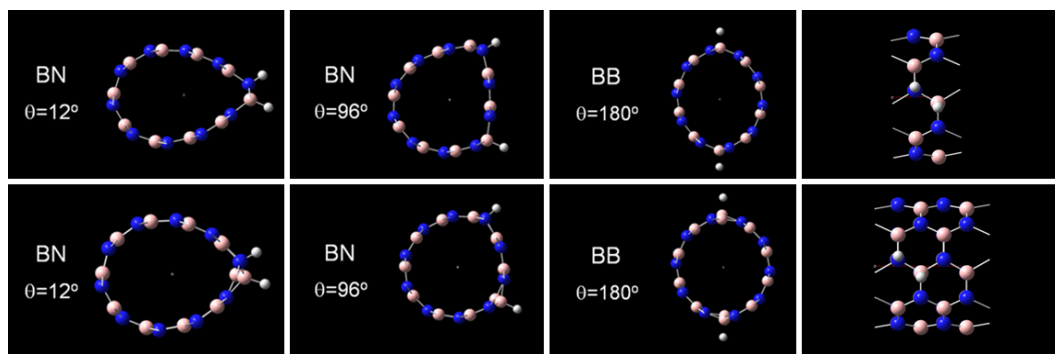

Figure 4: Demonstration of the BNNT cross-sectional deformation dependence on the inter-hydrogen angular separation for single (upper panels) and double (lower panels) unit-cells of the (5,5) BNNT. Geometry optimization of the presented structures was performed at the HSE/6-31G$^{**}$ level of theory. Side views of the single and double unit cells for the BN θ=12° adsorption scheme are presented in the rightmost panels.

Interestingly, for the ZZBNNTs considered, at some small hydrogen angular separations of the NN adsorption family, the high strain caused by the hydrogen adsorption results in cross-sectional buckling (see Fig. 5). Results obtained for this type of structural defect, which also appears when lowering the hydrogen density, are marked with dotted triangles in Figs. 2 and 6 and in supplementary Figs. S1-S4.

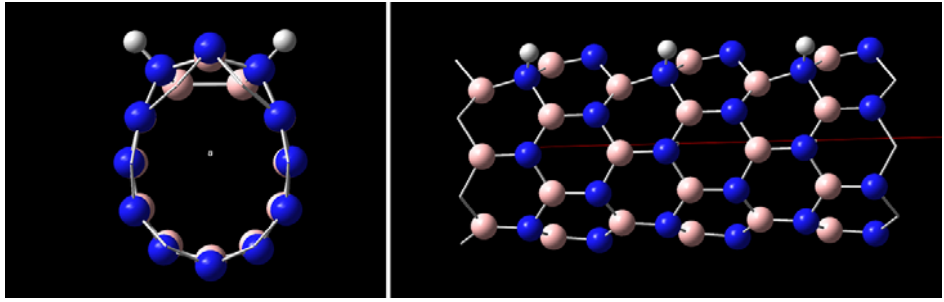

Figure 5: Axial (left panel) and side (right panel) views of the buckling structural defect occurring at the NN θ=60° position of the (6,0) ZZBNNT.

Electronic properties

We now turn to study the influence of hydrogen adsorption on the electronic properties of the BNNTs. We start by considering the effect of the adsorbate on the bandgaps of the various systems. In Fig. 6 we present the bandgap (calculated at the HSE/6-31G$^{**}$ level of theory) of the hydrogenated BNNTs as a function of the angular separation between the hydrogen adsorbates for the four BNNTs studied. Results obtained using the LDA and PBE functional approximations show similar trends and are presented in the supplementary material.

For the ACBNNTs studied we find that the decrease in bandgap from the pristine system to the most stable hydrogen adsorbed BNNT (smallest angular separation) is quite significant ~2.5 eV. As shown above, further increase of the hydrogen angular separation results in reduced energetic stability accompanied by a monotonous decrease in bandgap down to values of 0-0.2 eV for the BN family (apart from the (8,8) BNNT where for large angular separations a triplet ground state is obtained presenting a bandgap of ~1.5 eV). Interestingly, the BB and NN adsorption schemes, which are characterized by a triplet ground state, present a different behavior where the sensitivity of the bandgap towards the angular separation is considerably reduces. For both ACBNNTs studied the BB (NN) bandgaps are found to be in the range of 0.5-1.5 eV (4-5 eV).

For the ZZBNNTs studied a similar inverse relation is obtained between the hydrogen angular separation and the bandgap for the BN adsorption schemes. Nevertheless, the decrease in bandgap between the pristine and hydrogenated systems is much smaller (~0.2 eV) and the overall sensitivity of the bandgap with respect to the angular separation is lower thus small bandgap and metallic structures are obtained at much larger angles. Here, as well, the BB and NN adsorptions schemes present little dependence on the angular separation with bandgap values similar to those reported for the AC systems. When reducing the hydrogen density a qualitatively similar picture arises where the bandgap is inversely proportional to the hydrogen angular separation for the BN adsorption schemes though, as may be expected, the overall angular sensitivity of the bandgap is considerably reduced (see supplementary material for further details).

We therefore find that both the armchair and the zigzag BNNTs studied present a diverse range of bandgaps as a function of the hydrogen functionalization scheme and adsorbate density. This suggests that chemical decoration of BNNTs with hydrogen atoms can be used not only as a route for hydrogen storage but also as a means of tailoring the electronic properties of these systems.

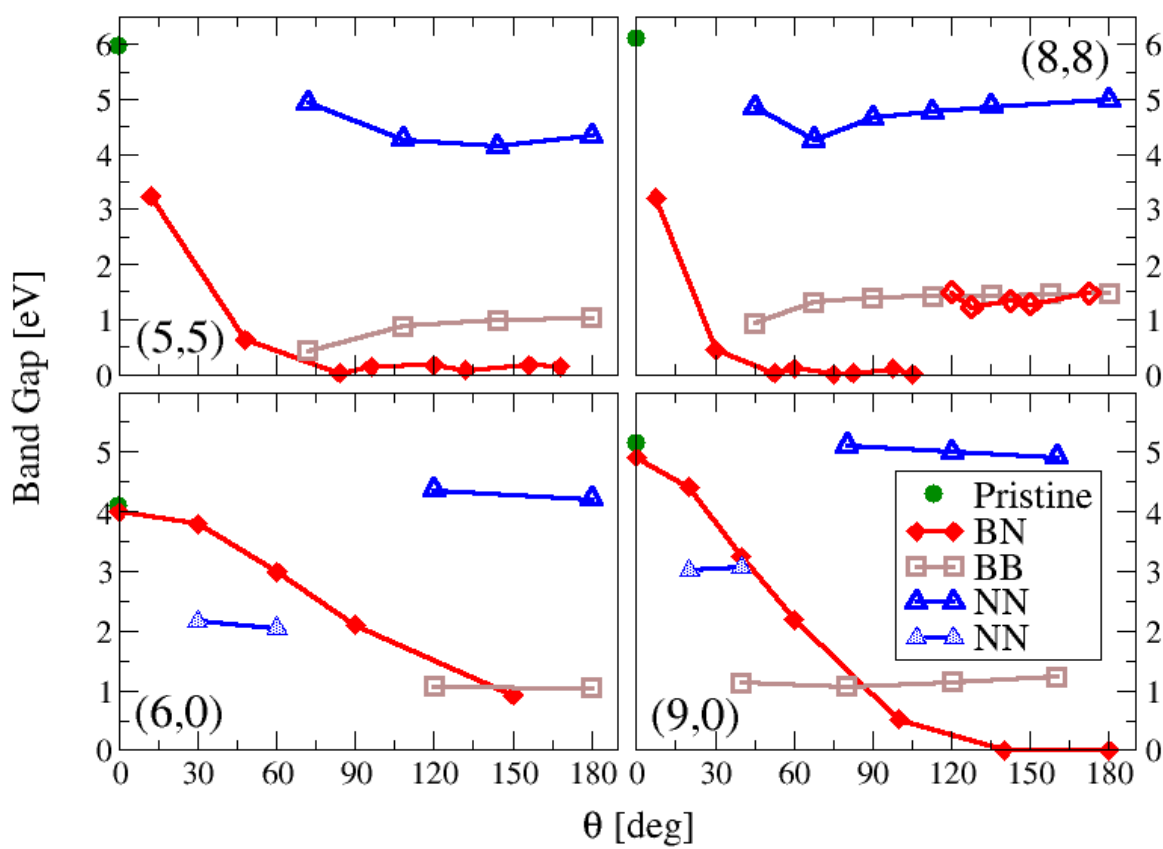

Figure 6: Band Gap as a function of inter-hydrogen angular separation (θ) for the (5,5) and (8,8) armchair (upper panels) and (6,0) and (9,0) zigzag (lower panels) BNNTs studied obtained at the HSE/6-31G$^{**}$ level of theory. For clarity, lines connecting values that belong to the BN (red diamonds), BB (brown squares), and NN (blue triangles) adsorption families were added. Full (open) symbols designate a closed-shell singlet (open-shell triplet) ground state. Dotted triangles represent NN adsorption schemes that form buckling deformations during geometry optimization (see Fig. 5 above and the relevant discussion).

To better understand these changes in the electronic properties we calculated the full band structure and partial density of states (PDOS) of the most stable hydrogen adsorption schemes of the armchair (Fig. 7) and zigzag (Fig. 9) BNNTs studied. Focusing first on the (5,5) armchair BNNT we find that the pristine (5,5) BNNT has an indirect band gap of 5.98 eV. From the PDOS analysis we find that the major contribution to the valence band comes from the nitrogen sub-network whereas the major contribution to the conduction band comes from the boron counterpart. This is expected in light of the differences in electronegativity between the boron and nitrogen atoms leading to partial charge transfer from the boron to the nitrogen sites. Upon hydrogen adsorption, at the lower density considered, the bandgap slightly reduces to 5.85 eV remaining indirect, the bands considerably flatten and the band degeneracies occurring in the pristine case are lifted.

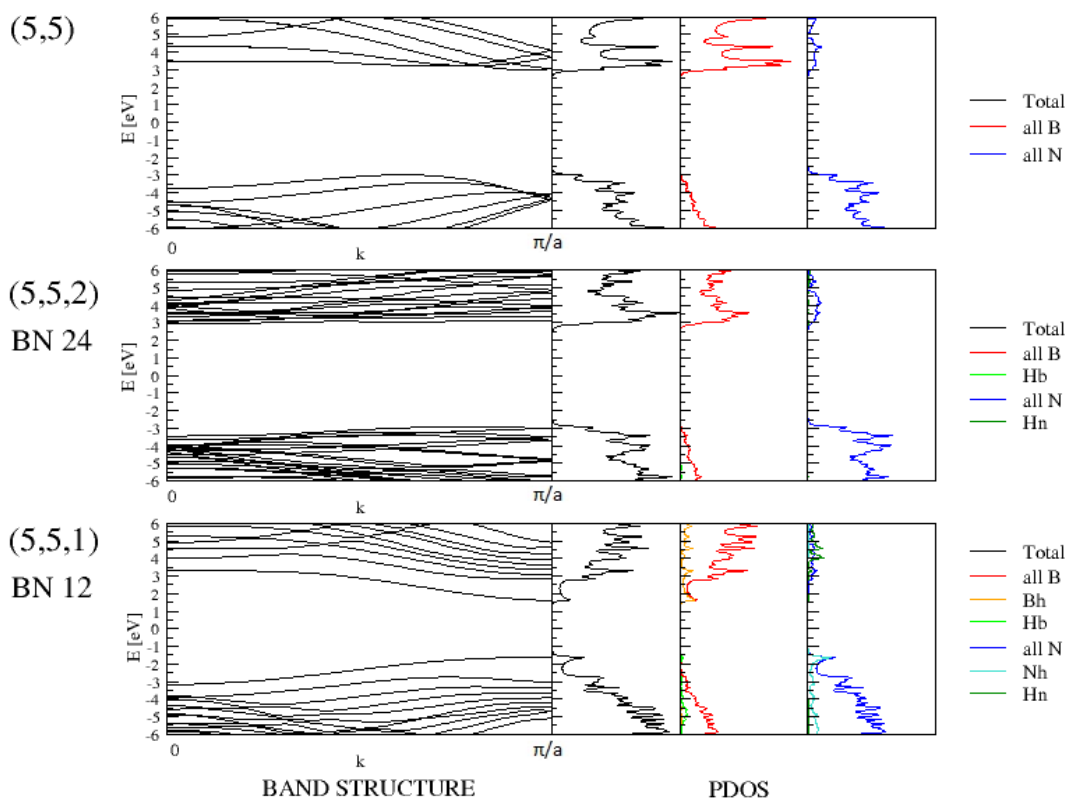

Figure 7: The band structures (left) and the PDOSs (right) obtained at the HSE/6-31G$^{**}$ level of theory for the pristine (5,5) ACBNNT (top panels), and the hydrogenated system at the lower (middle panels) and higher (bottom panels) hydrogen densities considered. For the hydrogenated systems results of the lowest energy configurations are shown. In the PDOS diagrams "Total" stands for total DOS; "all B" and "all N" stand for PDOS contribution of all the boron or all the nitrogen atoms, respectively; "$H_b$" and "$H_n$" stand for the PDOS contribution from the hydrogen atom bonded to a boron or nitrogen atom, respectively; and "$B_h$" ("$N_h$") stands for the PDOS contribution from boron (nitrogen) atom adjacent to the nitrogen (boron) adsorption site (see Fig. 8). The notations (5,5,1) and (5,5,2) stand for a hydrogen density of two hydrogen atoms per single BNNT unit-cell and two hydrogen atoms per two BNNT unit-cells, respectively.

When viewing the PDOS diagrams a picture similar to the pristine case arises where the valence (conduction) bands are constructed mainly from nitrogen (boron) atomic contributions with marginal contributions from the hydrogen atoms close to the Fermi energy. Interestingly, when increasing the hydrogen adsorption density a considerably smaller direct bandgap of 3.24 eV is predicted to occur accompanied by increased band dispersion. The PDOS analysis reveals that in this case considerable contributions both from the hydrogen adsorbates and from specific atomic positions adjacent to the adsorption sites are obtained. Specifically, the nitrogen atom adjacent to the boron adsorption site (see Fig. 8) is responsible for the van-Hove singularity obtained at the top of the Valence band whereas the boron atom adjacent to the nitrogen adsorption site (see Fig. 8) builds the corresponding DOS peak at the bottom of the conduction band. Similarly, the hydrogen atom positioned at the nitrogen adsorption site gives rise to large contributions to the conduction band DOS whereas its boron bonded counterpart contributes to the valence band DOS.

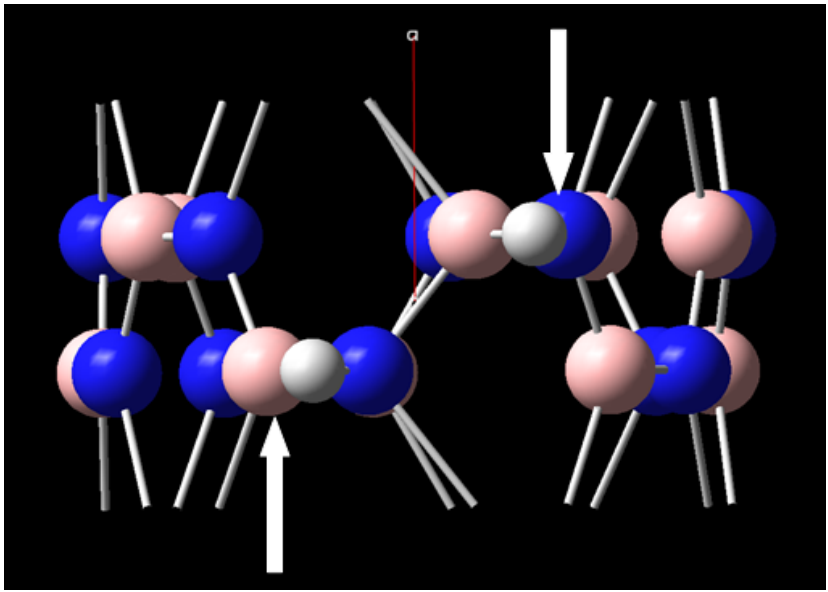

Figure 8: Demonstration of the atomic positions that have the most prominent contribution to the DOS of the hydrogenated (5,5) BNNT at the high adsorption density considered. The arrows mark the positions of the nitrogen (blue sphere) and boron (pink sphere) sites adjacent to the adsorption position, respectively.

Turning to discuss the zigzag system, Fig.9 presents the band structure and PDOS analysis for the (6,0) ZZBNNTs. The pristine system has a direct bandgap of 4.10 eV. Upon hydrogen adsorption, at the lower density considered, the bandgap remains direct and reduces to 3.95 eV. Similar to the armchair case, here as well there is a noticeable flattening of the bands and removal of the degeneracies existing in the pristine system. Nevertheless, from the PDOS we learn that the bands do not originate from individual contributions of specific atomic positions. Instead, similar to the case of the pristine nanotube, the valence bands are constructed mainly from the overall sum of all nitrogen atoms contributions and the conduction bands are mainly constructed from all boron atoms contributions. We note that the hydrogen atom positioned at the boron adsorption site gives rise to some DOS near the valence band maximum (see inset in the middle panel of Fig. 9). When considering the high hydrogen adsorption density scheme we notice that the bandgap changes only slightly to a value of 3.99 eV and remains direct with somewhat less band flattening. The PDOS analysis reveals a picture similar to the case of the lower adsorption density.

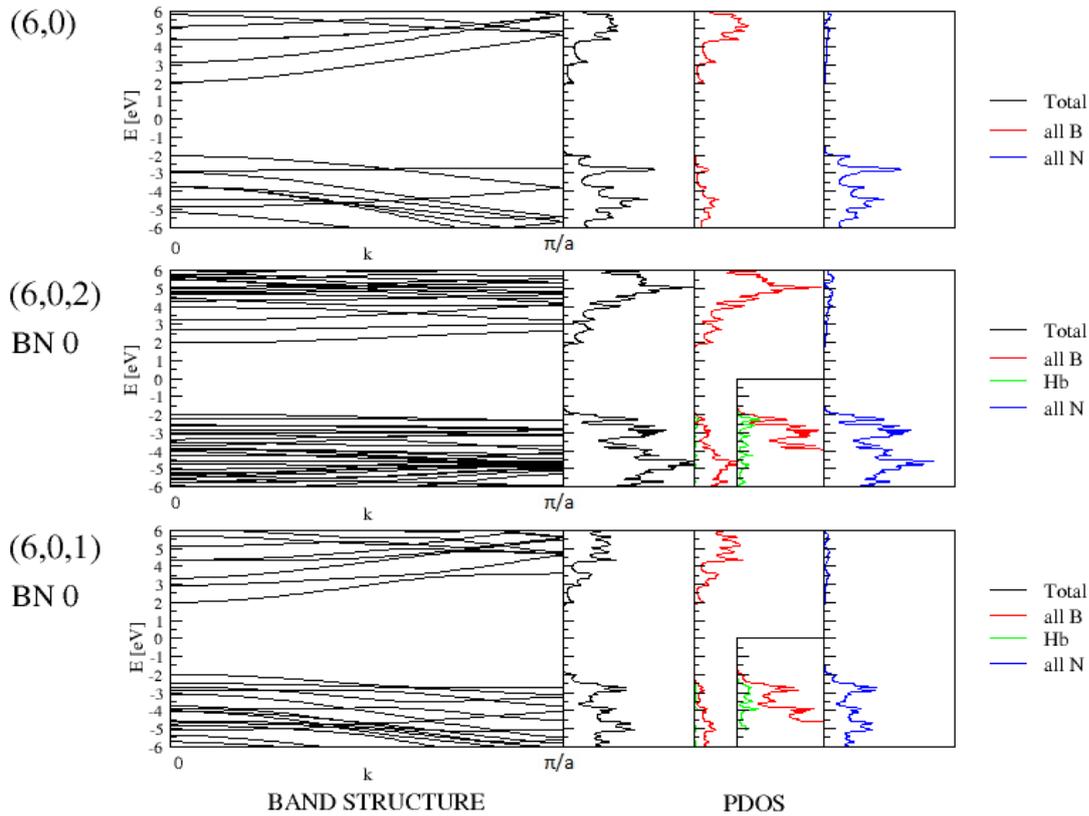

Figure 9: The band structures (left) and the PDOSs (right) obtained at the HSE/6-31G$^{**}$ level of theory for the pristine (6,0) ZZBNNT (top panels), and the hydrogenated system at the lower (middle panels) and higher (bottom panels) hydrogen densities considered. For the hydrogenated systems results of the lowest energy configurations are shown. Insets in the middle and lower panels present zoom-in on the density of states at the respective energy region. In the PDOS diagrams "Total" stands for total DOS; "all B" and "all N" stand for PDOS contribution of all the boron or all the nitrogen atoms, respectively; "$H_b$" stands for the PDOS contribution from the hydrogen atom bonded to a boron atom.

Finally, to get a clearer picture of the results presented in the band-structure analysis we visualize the valence band maximum (VBM) and conduction band minimum (CBM) orbitals of the most stable hydrogen adsorption schemes of the armchair (Fig. 10) and zigzag (Fig.11) BNNTs studied. The rotational symmetry of the pristine systems, which is responsible for the band degeneracies discussed above, is clearly reflected in the plotted orbitals. Furthermore, these orbitals are delocalized over the full nanotube circumference resulting in relatively high dispersion of the respective bands. Upon hydrogen adsorption, at the lower density considered, the symmetry is reduced and therefore the degeneracies are lifted. For the armchair systems considered we observe a strong localization of the orbitals around the hydrogen adsorption sites which explains the flattening of the corresponding bands. Increasing the hydrogen adsorption density leads to increased delocalization of the orbitals (compared to the low density) that results in an enhancement of the bands dispersion.

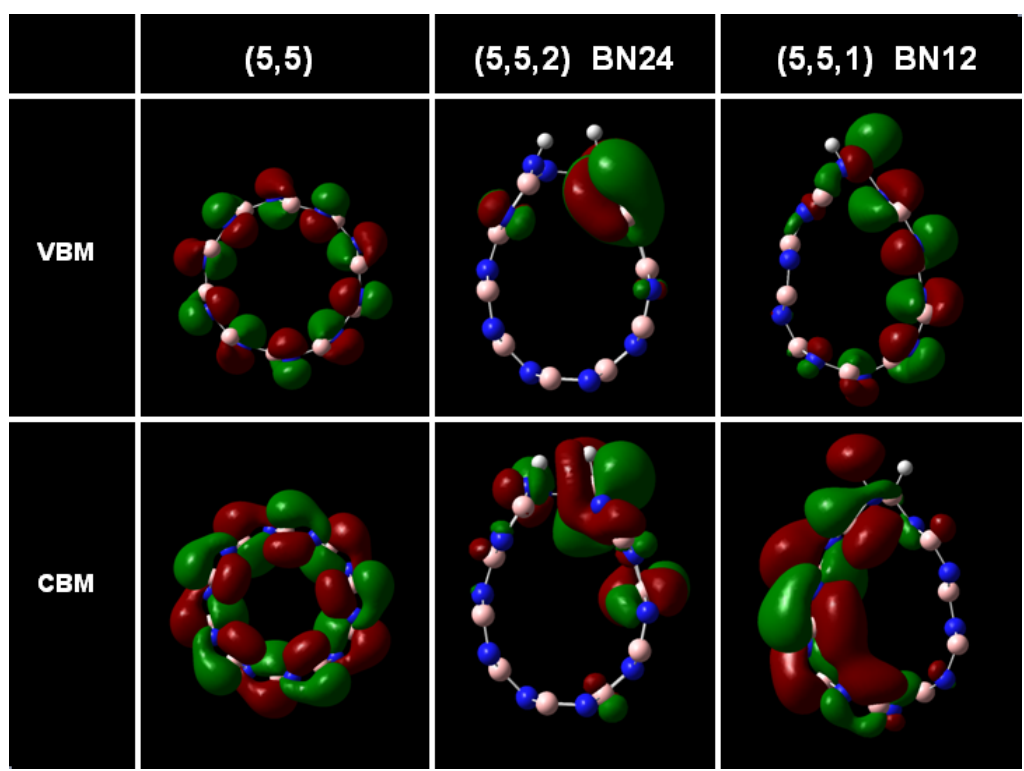

Figure 10: VBM (upper panels) and CBM (lower panels) of the pristine (second column), and hydrogen adsorbed (5,5) ACBNNT at the optimal hydrogen adsorption scheme of lower adsorption density (third column) and higher adsorption density (fourth column). Obtained at the HSE/6-31G** level of theory. Isosurface values are ±0.02 a.u.

For the ZZBNNTs studied a different picture is obtained. The VBM and CBM orbitals maintain their general character (in terms of nodal structure and symmetry) even upon hydrogen adsorption with no apparent localization. This results in much less flattening of the bands and minor variations in the bandgaps of the corresponding systems. We attribute these differences between the armchair and zigzag systems to the fact that the optimal hydrogen adsorption scheme of the zigzag system is of axial nature ($\theta=0^o$) whereas the optimal configuration for the armchair nanotube has a circumferential component ($\theta\neq0^o$). This results in higher symmetry of the zigzag adsorbed system which resembles more its pristine counterpart.

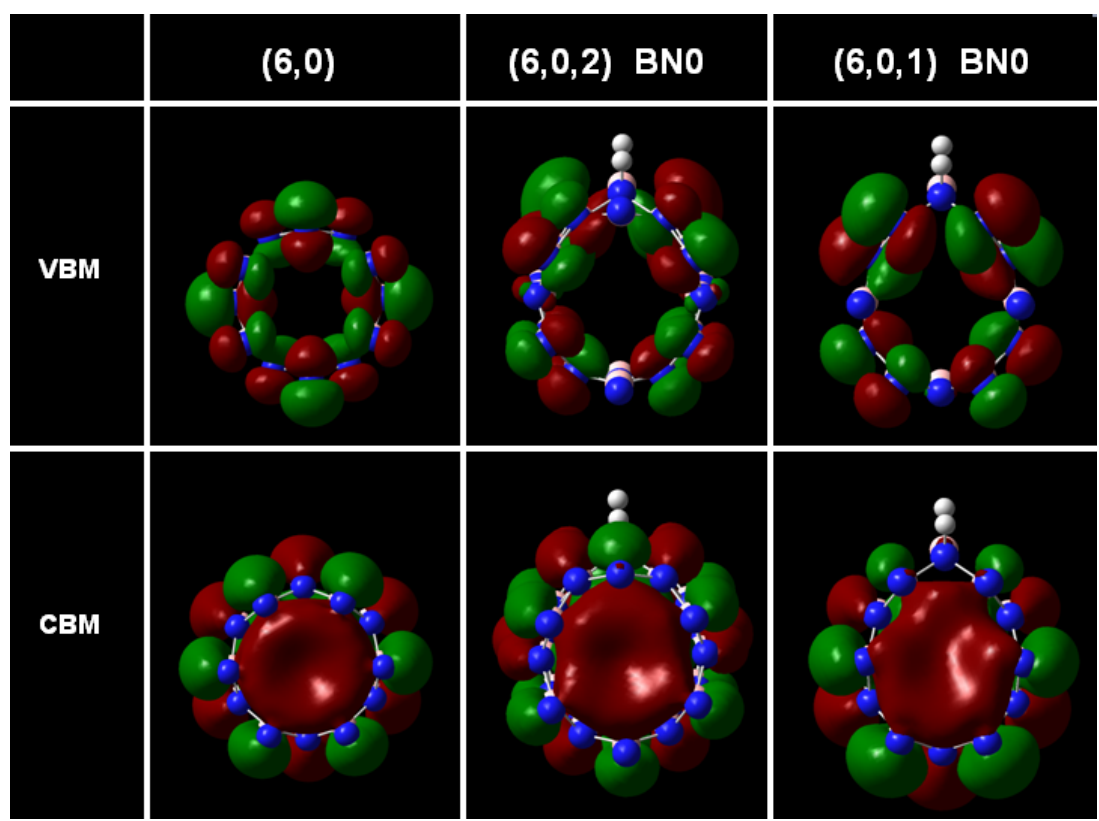

Figure 11: VBM (upper panels) and CBM (lower panels) of the pristine (second column), and hydrogen adsorbed (6,0) ZZBNNT at the optimal hydrogen adsorption scheme of lower adsorption density (third column) and higher adsorption density (fourth column). Obtained at the HSE/6-31G** level of theory. Isosurface values are ±0.02 a.u.

## Summary and Conclusions

In summary, we have investigated the effects of partial hydrogenation on the structural and electronic properties of boron nitride nanotubes. For the armchair and zigzag nanotubes studied the most energetically stable adsorption scheme was found to be that where one hydrogen atoms resides on a boron site and a second hydrogen atom resides on the adjacent nitrogen site. For small angular separations between two hydrogen adsorption sites the nanotubes undergo strong structural deformations where a transition from a circular to a "tear-like" cross section occurs. Furthermore, when the two hydrogen atoms reside on two nitrogen sites cross-sectional buckling is evident. As the angular separation increases the overall strain seems to decrease and the structure forms two arc-shaped segments of different curvatures which even up at $180^o$. The electronic properties of the BNNTs are found to be highly sensitive to partial hydrogenation where the bandgap decreases gradually upon increase of the angular separation between the two hydrogen adsorption sites. Furthermore, this sensitivity has unique characteristics depending on the nanotube identity, its diameter, the adsorbate density, and the chemical nature of the adsorption sites. For the armchair nanotubes studied a considerable reduction of the bandgap of the pristine BNNT is found upon hydrogen adsorption at the optimal adsorption scheme for the higher adsorption density considered (~2.5 eV). On the contrary, when considering the zigzag BNNTs, the optimal adsorption scheme is of axial nature leading to a preservation of the general character of the CBM and VBM and hence a considerably smaller reduction of the bandgap (in the order of 0.2 eV). These findings suggest that partial hydrogenation may become a valuable control tool for tailoring the electronic properties of boron nitride nanotubes and changing their insulating character into semiconducting and even metallic thus opening the way for diverse applications in the fields of nanoelectromechanical systems and nanoelectronics.

## Acknowledgments

This work was supported by the Israel Science Foundation under Grant No. 1313/08, the Center for Nanoscience and Nanotechnology at Tel Aviv University, the Israeli Ministry of Defense, and the Lise Meitner-Minerva Center for Computational Quantum Chemistry.